\documentclass[10pt, conference, letterpaper]{IEEEtran}
\IEEEoverridecommandlockouts
\usepackage{cite}
\usepackage{amsmath,amssymb,amsfonts}
\usepackage{algorithmic}
\usepackage{graphicx}
\usepackage{textcomp}
\usepackage{xcolor}

\usepackage{booktabs}
\usepackage{algorithm} 
\usepackage{subfigure}
\usepackage{balance}
\usepackage[misc]{ifsym}

\def\BibTeX{{\rm B\kern-.05em{\sc i\kern-.025em b}\kern-.08em
    T\kern-.1667em\lower.7ex\hbox{E}\kern-.125emX}}
\begin{document}

\title{6GAN: IPv6 Multi-Pattern Target Generation via Generative Adversarial Nets with Reinforcement Learning
}

\author{\IEEEauthorblockN{Tianyu Cui$^{1,2}$, Gaopeng Gou(\Letter)$^{1,2}$,  Gang Xiong$^{1,2}$, Chang Liu$^{1,2}$, Peipei Fu$^{1,2}$ and Zhen Li$^{1,2}$}
\IEEEauthorblockA{1.Institute of Information Engineering, Chinese Academy of Sciences\\
2.School of Cyber Security, University of Chinese Academy of Sciences\\
 \{cuitianyu, gougaopeng, xionggang, liuchang, fupeipei, lizhen\}@iie.ac.cn}}

\maketitle

\begin{abstract}
Global IPv6 scanning has always been a challenge for researchers because of the limited network speed and computational power. Target generation algorithms are recently proposed to overcome the problem for Internet assessments by predicting a candidate set to scan. However, IPv6 custom address configuration emerges diverse addressing patterns discouraging algorithmic inference. Widespread IPv6 alias could also mislead the algorithm to discover aliased regions rather than valid host targets. In this paper, we introduce 6GAN, a novel architecture built with Generative Adversarial Net (GAN) and reinforcement learning for multi-pattern target generation. 6GAN forces multiple generators to train with a multi-class discriminator and an alias detector to generate non-aliased active targets with different addressing pattern types. The rewards from the discriminator and the alias detector help supervise the address sequence decision-making process. After adversarial training, 6GAN's generators could keep a strong imitating ability for each pattern and 6GAN's discriminator obtains outstanding pattern discrimination ability with a 0.966 accuracy. Experiments indicate that our work outperformed the state-of-the-art target generation algorithms by reaching a higher-quality candidate set.
\end{abstract}

\begin{IEEEkeywords}
IPv6, Internet-wide Scanning, Network Measurement, Deep Learning, Reinforcement Learning
\end{IEEEkeywords}

\section{Introduction}
The exploration of the global IPv6 address space \cite{deering2017internet} could enhance the ability of researchers to conduct wide-ranging assessments of the next-generation Internet. Except for the particular case \cite{fiebig2017something}, the concrete instantiation of prior exploration work mainly includes passive measurement \cite{plonka2015temporal,czyz2014measuring,czyz2013understanding} and active scanning \cite{gasser2016scanning,beverly2018ip,gasser2018clusters,rye2020discovering,czyz2016don}. Because the former approach is limited by vantage points to monitor the traffic, fast IPv6 scanning has become a critical means required by the community.

However, IPv6 active hosts distribute in an extensive address space. A brute-force scanning approach \cite{durumeric2013zmap} with current network speed and computational power will expanse considerable time, rendering exhaustive probing completely infeasible. The recently proposed target generation algorithms \cite{ullrich2015reconnaissance,foremski2016entropy,murdock2017target,liu20196tree,cui20206gcvae,cui20206veclm} seem the state-of-the-art breakthrough to overcome the problem. Algorithms are required to learn features of the seed set and predict the active individuals or regions in the real network space to provide the candidate set waiting for scanning. The algorithmic design becomes essential to guarantee the quality of the candidate set.

While prior work has developed sophisticated techniques, IPv6 Internet nature begets these algorithms still facing challenges: \textbf{(1) IPv6 addressing pattern.} Network administrators are allowed to freely select IPv6 address configuration schemes \cite{hinden2006ip}, which enables multiple allocation pattern for interface identifier (IID) in the address. Clients may use Stateless Address Autoconfiguration and thus produce pseudorandom \cite{narten2001privacy} or EUI-64 IIDs \cite{thomson1998ipv6}, while servers and routers are usually assigned according to administrator customs or employing DHCPv6 \cite{bound2003rfc3315}. The algorithmic inference will be stuck in trouble as these patterns are required to be opaque according to RFC 7136 \cite{carpenter2014significance}. \textbf{(2) IPv6 aliasing.} Prior experience \cite{murdock2017target} has proposed that the large-scale aliased regions are the challenge that must be resolved in future IPv6 scanning because these addresses unconditionally respond to queries and seem not bound to unique devices. While the related alias detection work \cite{gasser2018clusters,liu20196tree} has gradually matured, these approaches are all scanning strategies for generating candidate sets or regions rather than algorithmic optimization. Therefore, the algorithm is still deceived to learn the aliased regions and consume a large budget to generate low-quality candidate sets. 

To address these problems, we raise a novel consideration to help push the candidate set from quantity to quality. Firstly, with the application of unsupervised learning to address sequences \cite{gasser2018clusters,cui20206veclm}, addressing patterns could be clustered into limited categories. We followed the idea to perform a deeper analysis under each pattern rather than the entire seed set to avoid multiple pattern interference. Secondly, in the field of natural language processing, the policy gradient of reinforcement learning \cite{sutton2000policy,williams1992simple} helps Generative Adversarial Nets (GANs) \cite{goodfellow2014generative} solve the non-differentiable problem of the discrete text. The generator is guided by the discriminator rewards and keeps improving with sampling policy just like an agent. We leverage this design to generate address sequences and increase a new reward mechanism with more intelligence to discourage aliased address generation.

\begin{table*}[htbp]
\caption{Comparison of Target Generation Algorithms}
\begin{center}
\begin{tabular}{llll}
\toprule
Approach&Target Generation&Alias Detection&Goal\\
\midrule
Entropy/IP \cite{foremski2016entropy} &Analyzing addressing structures through information entropy&-&Visual address distribution\\
6Gen \cite{murdock2017target} &Searching the densest address clusters to provide active regions&Sampling scanning &Remarkable performance\\
6Tree \cite{liu20196tree} &Dynamic adjusting search directions with a space tree&Dynamic scanning &Faster time complexity\\
6GCVAE \cite{cui20206gcvae} &Reconstructing addresses through variational autoencoder&-&Deep learning attempts\\
6VecLM \cite{cui20206veclm} &Predicting address sequences through language modeling&-&IPv6 semantics exploration\\
\textbf{6GAN}&\textbf{Multi-pattern target generation through adversarial training} &\textbf{Reinforcement learning}&\textbf{Higher-quality candidates}\\
\bottomrule
\end{tabular}
\label{tab1}
\end{center}
\end{table*}

In this paper, we introduce a deep learning architecture 6GAN to replace traditional target generation algorithms. 6GAN uses multiple generators to learn IPv6 addressing patterns determined by known seed classification methods. These generators employ policy gradient \cite{sutton2000policy} and Monte Carlo search \cite{browne2012survey} for the address sequence decision-making process, which is motivated by two environmental rewards. One reward comes from a multi-class discriminator's estimation of whether the generated address conforms to the real pattern address, the other reward is the evaluation of the aliased prefix detection to judge whether to update the address prefix. The approach contributes to uncovering active targets under multiple addressing patterns and avoiding learning from aliased addresses. 

\textbf{Contributions.} Our contributions can be summarized as follows:
\begin{itemize}
\item We propose a novel architecture 6GAN to generate diversified non-aliased active addresses of different addressing pattern types through using multiple generators guided by rewards from a discriminator and an alias detector.
\item We employ a multi-class objective of 6GAN's discriminator, which can identify IPv6 addressing pattern categories.
\item We implement an alias detection approach embedded in the algorithm by optimizing the generator, which saves algorithmic budget to generate high-quality candidates.
\item We push the quality of candidate sets to a higher level. Experiments show that 6GAN outperforms state-of-the-art target generation algorithms on multiple metrics.
\end{itemize}

\textbf{Roadmap.} Section \ref{sec2} summarizes the prior researches related to our work. Section \ref{sec3} introduces the basic knowledge about IPv6 target generation. Section \ref{sec4} highlights the overall design of 6GAN. Section \ref{sec5} shows the evaluation results and Section \ref{sec6} concludes the paper.

\section{Related Work}\label{sec2}
Prior work on IPv6 target generation falls into two broad categories: (1) analyzing seed addresses to understand allocation patterns and (2) designing algorithms that generate candidate targets to scan. In addition, we will introduce (3) the related applications using GANs and reinforcement learning.

\subsection{Addressing Pattern Learning}
RFC 7707 \cite{chown2016network} points out known address configuration schemes and possible administrator configuration customs to facilitate the exploration of IPv6 address space. The measurement work in the document indicates that most addresses follow specific patterns. Gasser et al. \cite{gasser2018clusters} employed entropy clustering to classify the known addresses into 6 addressing pattern categories. The classification results show a great correlation with the configuration schemes. Cui et al. \cite{cui20206veclm} used IPv62Vec to explore IPv6 semantics to learn address similarity. Addresses with similar semantic structures are gathered in the same cluster in the vector space. These prior works determined the possible addressing patterns in the IPv6 space and laid the foundation for the target generation in addressing patterns.

\subsection{Target Generation Algorithms}
 \textbf{1) Traditional Design Algorithms.} Traditional design algorithms refer to target generation algorithms designed based on human experience and assumptions. 
\begin{itemize}
\item Ullrich et al. \cite{ullrich2015reconnaissance} used a recursive algorithm for the first attempt to target generation. The algorithm greedily determines the bit by reaching the most addresses in the seed set in each recursion until remaining reserved bits.
\item Foremski et al. \cite{foremski2016entropy} proposed Entropy/IP, which employs a Bayesian network to model the statistical dependence between the values of different entropy segments. The statistical model could generate addresses for scanning.
\item Murdock et al. \cite{murdock2017target} introduced 6Gen, which generates the densest address range clusters by combining the closest Hamming distance addresses in each iteration. 
\item Zhizhu et al. \cite{liu20196tree} implemented 6Tree, which performs information entropy to construct a space tree to discover active regions and dynamically adjust the exploration direction according to the scan results. The work achieves a faster linear time complexity compared to 6Gen.
\end{itemize}

 \textbf{2) Deep Learning Approaches.} The exploitation of deep learning approaches to achieve the algorithm has recently been developed because of the adaptability to different datasets.
\begin{itemize}
\item Cui et al. \cite{cui20206gcvae} proposed 6GCVAE to complete the first attempt to employ deep learning architecture. They used the Variational Autoencoder \cite{kingma2013auto} with gated convolutional layers to discover active addresses and proved that seed classification can improve the experimental performance.
\item Cui et al. \cite{cui20206veclm} introduced 6VecLM, which utilizes a Transformer network \cite{vaswani2017attention} to build an IPv6 language model, i.e., learning sequence relationships and using softmax temperature \cite{muller2019does} to generate creative addresses.
\end{itemize}

Table \ref{tab1} compares the instantiations of these works with our approach. 6GAN achieves address discovery in multiple patterns and detecting aliased regions during model training to generate high-quality candidate sets without wasting budget.

\subsection{Generative Adversarial Nets and Reinforcement Learning}
Generative Adversarial Nets (GANs) \cite{goodfellow2014generative} are a novel class of deep generative models with great success in computer vision. However, discrete sequences are not differentiable, causing little progress in text generation. Yu et al. \cite{yu2017seqgan} leveraged reinforcement learning which treats the process of discriminator guiding generator as a reinforcement learning policy \cite{sutton2000policy,williams1992simple} to tackle the problem and led more GANs to follow this scheme in natural language generation. Wang et al. \cite{wang2018sentigan} proposed SentiGAN to generate text with multiple sentiment labels, which is similar to our purpose. While we aim to perform IPv6 multi-pattern target generation and employ two rewards to co-instruct generators to complete active address generation and alias detection.

\section{Preliminaries}\label{sec3}
In this section, we provide the target generation problem definition and a brief IPv6 background related to our work to help readers understand this paper.

\subsection{Problem Definition}
The IPv6 target generation problem requires an algorithm $\tau$ fed with a seed set $S$ to predict a possible active candidate set $C$ with a limited budget $|C|$. Considering the influence of aliased addresses, we assume that $T$ is the set including all active addresses in the IPv6 space and $T_a$ is a subset of $T$ including all aliased addresses. The algorithm $\tau$ finally discovers a valid target set $\widehat{C}$:
\begin{equation}
\widehat{C} = C\cap T - C\cap T_a - C\cap S
\end{equation}
Where $C\cap T$ could be verified through a scanner like Zmapv6 \cite{durumeric2013zmap}. $C\cap T_a$ could be obtained by alias detection. $C\cap S$ is collected through algorithmic search.

To ensure a high-quality candidate set, the algorithm $\tau$ requires minimizing the objective function $L_{\tau}$ to reach optimal:
\begin{equation}
L_{\tau} = |C| - |\widehat{C}|
\end{equation}

\subsection{IPv6 Background}\label{sec3.2}
 \textbf{1) IPv6 Addressing.} An IPv6 address consists of a global routing prefix, a local subnet identifier, and an interface identifier (IID) \cite{hinden2006ip}. While the global routing prefix is determined to route traffic destined to a Local Area Network (LAN), the configuration of IID is allowed more freedom to ensure the uniqueness of the host interface in the local network segment. According to RFC 7707 \cite{chown2016network}, these IIDs could be categorized into the following patterns:
\begin{itemize}
\item \textbf{Embedded-IPv4.} In which the lowest-order 32 bits or each byte of IID embeds IPv4 address of the network interface. For instance, 0:0:c0a8:20a or c0:a8:2:a could embed an IPv4 address 192.168.2.10.
\item \textbf{Embedded-port.} In which the lowest-order byte of the IID embeds the service port and the rest bytes are all zero. For instance, 0:0:0:80 or 0:0:0:50 embed decimal or hexadecimal port 80 for HTTP.
\item \textbf{IEEE-derived.} In which the word “fffe” is inserted between the Organizationally Unique Identifier (OUI) \cite{narten2001privacy} and the rest of the Ethernet address. For instance, 0250:56ff:fe89:49be.
\item \textbf{Low-byte.} In which all the bytes of the IID are set to zero except the least significant bytes in one or two lowest-order. For instance, 0:0:0:a or 0:0:1:a.
\item \textbf{Pattern-bytes.} In which more than two bytes of the IID have identical values and belong to the same prefix. These IIDs contain a specific addressing pattern different from the above.
\item \textbf{Randomized.} In which the IID keeps a pseudorandom representation because these addresses might be privacy addresses by using Stateless Address Autoconfiguration (SLAAC) \cite{thomson1998ipv6}.
\end{itemize}
The complex address composition raises the significance of classification. Target generation algorithms are urgently required deep eyes on each addressing pattern rather than bearing the pressure of the whole IPv6 address space.

 \textbf{2) Aliased Address.} Aliased addresses refer to all addresses under aliased prefixes, which unconditionally respond to scan queries. For instance, the address 2001:db8::20:1a could be recognized as an aliased address because of the known aliased prefix 2001:db8::/32. These aliased regions could be formed through configuration like directed mapping to the same server. Prior work \cite{murdock2017target} indicates more appearance in large CDN networks. Aliased addresses seriously affect the accuracy of host discovery approaches because there is no mapping relationship between these addresses and unique devices. Performing alias detection has been a consensus in IPv6 scanning.
 
However, previous work \cite{murdock2017target,gasser2018clusters} detected the aliased address mainly on the candidate set and ignored that algorithms still consume the budget to generate aliased addresses. These aliased addresses mislead the algorithmic inference during algorithm learning. Even though alias detection is pre-processed on the seed set, algorithms may reconstruct the alias address during prediction through sampling strategy. Therefore, an advanced algorithm must consider how to discourage learning aliased prefixes during algorithmic execution.

\begin{figure*}[htbp]
\centerline{\includegraphics[width=0.95\textwidth]{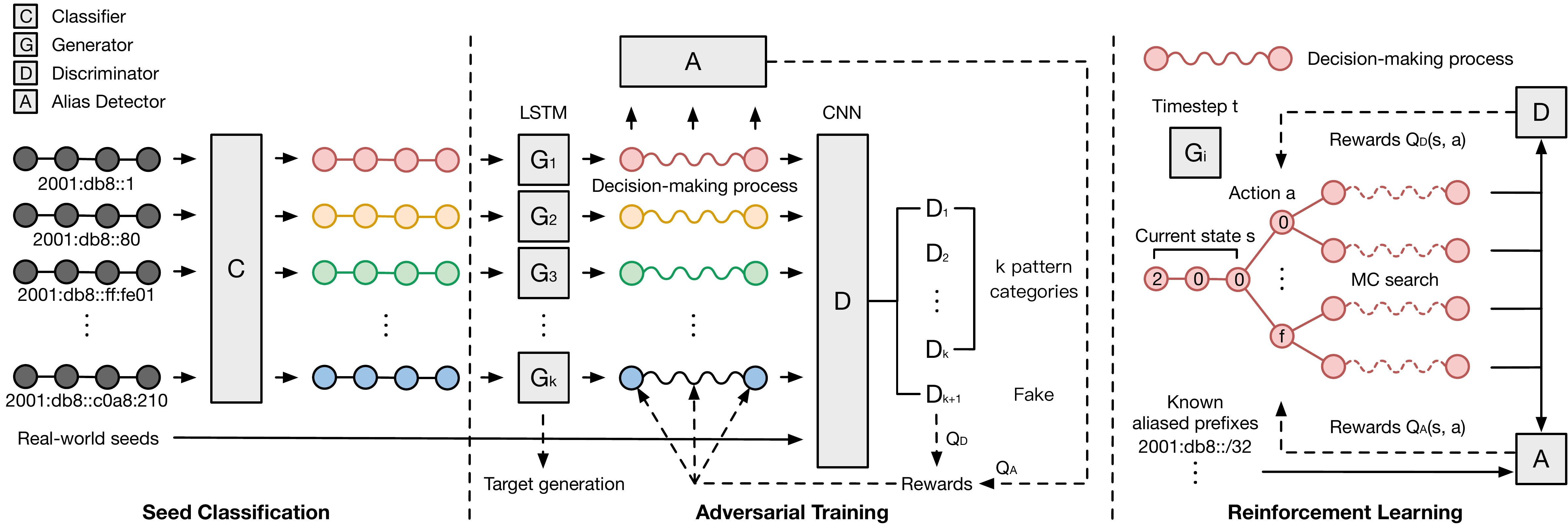}}
\caption{The overall architecture of 6GAN. Real-world seed addresses are classified into $k$ addressing patterns in seed classification. $k$ generators imitate these pattern types through adversarial training with a multi-class discriminator. The address sequence decision-making process of the generators is indicated by the rewards from the discriminator and an alias detector through reinforcement learning during training.}
\label{fig1}
\end{figure*}


\section{6GAN}\label{sec4}
6GAN is a target generation architecture that integrates seed classification and adversarial training with reinforcement learning, which is shown in Figure \ref{fig1}. The architecture could be divided into four objectives: seed classification, generator learning, discriminator learning, and alias detection. 

The real-world seed addresses will first be performed pattern discovery through known seed classification methods. Then, $k$ generators will be respectively fed with the addresses of the $k$ classified addressing patterns and complete the adversarial training with a discriminator. The goal of the $i$-th generator $G_i$ is to generate addresses with the $i$-th pattern type to deceive the discriminator $D$, while the goal of the discriminator $D$ is to distinguish between fake addresses generated by generators and real addresses with $k$ pattern labels. The learning of generator $G_i$ is guided by the evaluation rewards of generated addresses, which is from the discriminator $D$ and an alias detector $A$ through reinforcement learning. The alias detector $A$ is fed with known aliased prefixes to estimate rewards. 

\subsection{Seed Classification}
To determine the hidden IPv6 addressing pattern in the real world, we classify the seed set and determine the number of generators $k$ according to the classification result. Based on previous experience, we provide the following three seed classification methods to promote pattern discovery:
\begin{itemize}
\item \textbf{RFC Based.} According to possible IPv6 addressing patterns proposed in RFC 7707 \cite{chown2016network}, the addr6 tool in ipv6toolkit \cite{gont2012security} can match the patterns mentioned in Section \ref{sec3.2}. We use the tool to perform the match on the seed set to obtain RFC-based classification results.
\item \textbf{Entropy Clustering.} Gasser et al. \cite{gasser2018clusters} proposed entropy clustering, which uses information entropy of the nybble value under the same prefix in the seed set as a prefix fingerprint and utilizes k-means \cite{jain2010data} to perform unsupervised clustering to discover the prefix-level pattern set. The hyperparameter $k$ controls the number of generated pattern categories. We label an address according to the pattern category to which the prefix corresponding to the address belongs.
\item \textbf{IPv62Vec.} Cui et al. \cite{cui20206veclm} proposed IPv62Vec, which implements the mapping from address space to vector space by learning the context of words in the address. IPv6 addresses with similar semantic structures are distributed in the same cluster in the vector space. The method finally employs DBSCAN \cite{ester1996density} to find these clusters to identify possible addressing patterns. We restore the category of the address corresponding to each vector by using the mapping relationship between addresses and vectors.
\end{itemize}
Where the first method is the address division based on measurement experience. The other two methods employ machine learning approaches and could discover more unknown addressing patterns. 6GAN could freely choose one of the classification methods to determine the addressing patterns to be learned by generators.

\subsection{Generator Learning}
In our model architecture, the target generation problem could be considered as an address sequence decision-making problem. A hexadecimal IPv6 address can be denoted as an address sequence $X_{0:T} = (x_0, ... , x_t, ..., x_T), x_t \in V = \{0, 1, ..., f\}$, where $V$ is the vocabulary including all address nybble values to be selected. When training a $\theta_i$-parameterized generator $G_{\theta_i}$ at timestep $t$, the state $s$ is the current produced address nybbles $X_{0:t-1} = (x_0, ..., x_{t-1})$ and the action $a$ is the next nybble value $x_t$ to be select. Following Sutton et al. \cite{sutton2000policy}, the policy $G_{\theta_i}(a|s)$ is the probability $G_{\theta_i}(a = x_t | s = X_{0:t-1})$ that action selects $x_t$ as the next nybble according to the previous sequence of the current state $X_{0:t-1}$. The action-value function $Q_{AD_\phi}^{G_{\theta_i}}(s, a)$ is the assessment $Q_{AD_\phi}^{G_{\theta_i}}(s = X_{0:t-1}, a = x_t)$ of the sequence $X_{0:t}$ led by the action policy based on the rewards of a $\phi$-parameterized discriminator $D_\phi$ and an alias detector $A$. Therefore, the objective function $J(\theta_i)$ of the $i$-th generator $G_{\theta_i}$ is:
\begin{equation}\label{eq3}
J(\theta_i) = \sum_{t=1}^TG_{\theta_i}(x_t|X_{0:t-1})Q_{AD_\phi}^{G_{\theta_i}}(X_{0:t-1}, x_t)
\end{equation}
Where $Q_{AD_\phi}^{G_{\theta_i}}(s, a)$ is a penalty based action-value function that we follow Wang et al. \cite{wang2018sentigan}. A high action value $Q_{AD_\phi}^{G_{\theta_i}}(s, a)$ could be considered that the generated address is recognized as a fake address by the discriminator or an aliased address marked by the alias detector. This reminds the generator to update the current policy to exploit other nybble values $x_t$ until the objective function $J(\theta_i)$ is minimized.

The action value $Q_{AD_\phi}^{G_{\theta_i}}(s, a)$ in 6GAN is composed of discriminator reward $Q_{D_\phi}^{G_{\theta_i}}(s, a)$ and alias detection reward $Q_A^{G_{\theta_i}}(s, a)$:
\begin{equation}\label{eq4}
Q_{AD_\phi}^{G_{\theta_i}}(s, a) = Q_{D_\phi}^{G_{\theta_i}}(s, a) + \alpha Q_A^{G_{\theta_i}}(s, a)
\end{equation}
Where $\alpha$ is the hyperparameter that controls the ratio of the two rewards.

At each timestep $t$, the discriminator and alias detector are required to evaluate the generator policy to obtain $Q_{AD_\phi}^{G_{\theta_i}}(s, a)$, while the generator only keeps the incomplete address sequence $X_{0:t}$ composed of the current state $X_{0:t-1}$ and the action $x_t$. To produce a complete sequence $X_{0:T}$ for judgment, the generator applies a $N$-time Monte Carlo search \cite{browne2012survey} with a roll-out policy $G_{\theta_i}$ to sample the future sequence $X_{t+1:T}$:
\begin{equation}
\text{MC}^{G_{\theta_i}}(X_{0:t}; N) = \{X_{0:T}^1, ..., X_{0:T}^N\}
\end{equation}
Where $N$ is the sampling times to produce a batch of address samples to more accurately calculate an assessment of the action value.

All generators of 6GAN use Long Short-Term Memory (LSTM) \cite{hochreiter1997long} cells to model $G_{\theta_i}(a|s)$. LSTM is a recurrent neural network outputs the $t$-th address nybble $x_t$ based on the hidden $h_t$ retaining the previous states:
\begin{equation}
\begin{split}
p(x_t | X_{0:t-1}) = \text{softmax}(c + wh_t)\\
\text{where} \quad h_t = \text{LSTM}(h_{t-1}, x_{t-1})
\end{split}
\end{equation}
Where the parameters are a bias matrix $c$ and a weight matrix $w$. Softmax function achieves the selection probability of $x_t$. 

It is worth noting that each generator of 6GAN applies the above framework separately without parameter sharing. Each generator independently learns the addressing pattern to generate diversified addresses.

\subsection{Discriminator Learning}
Our discriminator uses a multi-class classification objective to help distinguish between the real active addresses with addressing pattern types and the generated addresses. The objective is trained by providing the real-world seed addresses and  the synthetic addresses generated from generators. When 6GAN sets $k$ generators for training, the discriminator classifies the input addresses into $k+1$ categories by producing a softmax probability distribution to obtain the class score $D_\phi(X) = (D_\phi^1(X), ..., D_\phi^{k+1}(X))$. The $i$-th of the first $k$ scores represents the probability of a sample being judged as the $i$-th pattern type address. The $(k+1)$-th score denotes the probability of a sample being classified into the fake address class. The objective function $J(\phi)$ of the discriminator is:
\begin{equation}\label{eq7}
J(\phi) =  - \sum_{i=1}^k\mathbb{E}_{X\sim p_i}[\text{log}D_\phi^i(X)] - \mathbb{E}_{X\sim G_{\theta}}[\text{log}D_\phi^{k+1}(X)]
\end{equation}
Where $p_i$ is real $i$-th type pattern addresses. The discriminator is required to minimize the objective function $J(\phi)$ to optimize discrimination ability.

At each timestep $t$, the discriminator provides the reward $Q_{D_\phi}^{G_{\theta_i}}(s,a)$ through evaluating the batches of address samples generated by generators with $N$-time Monte Carlo search:
\begin{equation}
\begin{split}
&Q_{D_\phi}^{G_{\theta_i}}(s = X_{0:t-1},a = x_t) = \\
&\left\{ 
\begin{aligned} 
&\textstyle\frac{1}{N}\sum\nolimits_{n=1}^N(1 - D_\phi^i(X_{0:T}^n)), X_{0:T}^n\in \text{MC}^{G_{\theta_i}}(X_{0:t};N) &t < T\\ 
&1 - D_\phi^i(X_{0:t})  &t = T 
\end{aligned} 
\right.
\end{split}
\end{equation}

\begin{algorithm}[htbp] 
\caption{The adversarial training process in 6GAN} 
\label{alg1} 
\begin{algorithmic}[1] 
\REQUIRE $k$ Generators $G_{\theta_i}$, $i$ = $(1, ..., k)$; Discriminator $D_\phi$; Alias detector $A$; Real-world address seed set with $k$ types of addressing pattern $S = \{S_1, ..., S_k\}$ 
\ENSURE Optimal generators $G_{\theta_i}$ and discriminator $D_\phi$
\STATE Initialize $G_{\theta_i}$, $D_\phi$ with random wights $\theta_i$, $\phi$
\STATE Pre-train $G_{\theta_i}$ on $S_i$
\STATE Generate fake addresses $F$ using $G_{\theta_i}$ for training $D_\phi$
\STATE Pre-train $D_\phi$ on $S\cup F = \{S_1, ..., S_k, F\}$
\REPEAT 
\FOR{g-steps}
\STATE Generate $k$-type fake sequences $X_{0:T}$ using $G_{\theta_i}$
\FOR{$t$ in $0\sim T$} 
\STATE Calculate $Q_{AD_\phi}^{G_{\theta_i}}(s = X_{0:t-1},a = x_t)$ by Eq. \ref{eq4}
\ENDFOR
\STATE Update generators $G_{\theta_i}$ by minimizing Eq. \ref{eq3}
\ENDFOR 
\FOR{d-steps} 
\STATE Generate $k$-type fake addresses $F$ using $G_{\theta_i}$
\STATE Update $D_\phi$ using $S\cup F$ by minimizing Eq. \ref{eq7}
\ENDFOR  
\UNTIL{6GAN converges}
\RETURN
\end{algorithmic} 
\end{algorithm}

The discriminator of 6GAN is implemented using Convolutional Neural Networks (CNN) \cite{kim2014convolutional} with multiple filters. Filters of different sizes could contribute to discovering multiple address sequence structures. This framework enhances the ability of the discriminator to discriminate different addressing patterns, which helps to reach optimal in adversarial training.

In 6GAN, $k$ generators and one discriminator will be trained alternately to achieve their respective goals. Algorithm~\ref{alg1} shows the full details of the adversarial training on the generators and the discriminator.

\subsection{Alias Detection}
According to prior studies, the discovery work of aliased prefixes has gradually matured. In the 6GAN model, we consider applying these known aliased prefixes at the algorithmic level to help prevent the generation of aliased addresses, which form the alias detector $A$ in the architecture.

Consider an aliased prefix $P_{0:L} = (p_0, ...,p_t, ..., p_L)$, where $ L\leq T$, a generator with Monte Carlo search generates an address sequence $X_{0:T}$. The alias detector identifies an aliased address when $P_{0:L} = X_{0:L}$. Therefore, the alias detector calculates an alias score $A(X)$ according to the judgment:
\begin{equation}
A(X) = 
\left\{
\begin{aligned} 
\lambda &&P_{0:L} = X_{0:L}\\
0 &&P_{0:L} \neq X_{0:L}
\end{aligned} 
\right.
\end{equation}
Where $\lambda$ is a hyperparameter to control the reward strength when an aliased address is detected.

At each timestep $t$, the alias detector evaluates generated the address sequence $X_{0:T}$ to provide the reward $Q_A^{G_{\theta_i}}(s, a)$: 
\begin{equation}
\begin{split}
&Q_A^{G_{\theta_i}}(s=X_{0:t-1}, a=x_t) =  \\
&\left\{ 
\begin{aligned}
&\textstyle\frac{t}{NL}\sum_{n=1}^{N}A(X_{0:T}^n), X_{0:T}^n\in\text{MC}^{G_{\theta_i}}(X_{0:t};N) &t\leq L\\
&0 & T \geq t >L
\end{aligned} 
\right.
\end{split}
\end{equation}
Where a positive reward $Q_A^{G_{\theta_i}}(s, a)$ is only provided on the prefix part. The coefficient $\frac{t}{L}$ performs hierarchical rewards for the address sequence $X_{0:L} = (x_0, .., x_t, ..., x_L)$. The high indexes of the prefix (like $x_L$) will obtain high rewards when meeting an aliased prefix. This design forces the generator to more likely update the high index and helps reduce the wide range changes of the prefix region.


\section{Evaluation}\label{sec5} 
In this section, we present our experimental setup and all experimental results to indicate 6GAN's performance. 

\subsection{Experimental Setup}
\textbf{1) Dataset.} The dataset used in the paper is collected from a public dataset IPv6 Hitlist \cite{gasser2018clusters} and a measurement dataset CERN IPv6 2018. Table \ref{tab2} shows the detail of these datasets.

\begin{itemize}
\item \textbf{IPv6 Hitlist.} Gasser et al. \cite{gasser2018clusters} provided the public dataset IPv6 Hitlist to help future IPv6 study. The dataset is formed by the results daily probing multiple public address sets. They also used scanning strategies to provide daily updated aliased prefix data. To ensure the reproducibility of the 6GAN’s performance, we use the address data as the real-world seed set and utilize the aliased prefix data as the benchmark for 6GAN's alias detector.
\item \textbf{CERN IPv6 2018.} To explore the performance difference of the model on different datasets, we additionally employ the passive measurement dataset CERN IPv6 2018 to indicate the 6GAN's robustness. We passively collected address sets on China Education and Research Network (CERNET) from March to July 2018 and provided keeping-active addresses to build the dataset.
\end{itemize}

\textbf{2) Verification Method.} To ensure the accuracy of the results, we design the scanning operation to verify the active addresses generated through target generation algorithms. Firstly, we use Zmapv6 tool \cite{durumeric2013zmap} to perform scanning work through multiple protocols including ICMPv6, TCP/80, TCP/443, UDP/53, and UDP/443. An address is considered an active target when any scanning protocol obtains a response. Secondly, to reduce the impact of the difference in host activity at different times, we maintain continuous scanning for three days. An address responding to the scanning on any one day will be considered an active target.

\textbf{3) Evaluation Metric.} To evaluate the quality of the generated candidate set, we design multiple evaluation metrics to measure from various aspects.
\begin{itemize}
\item \textbf{Pattern quality.} We propose the pattern quality metric to evaluate the imitating ability of the generators to each addressing pattern. Given a real-world address seed set with $k$ types of addressing pattern $S = \{S_1, ..., S_t, ..., S_k\}$, a candidate set $C$ is generated by using the $t$-th pattern generator. The pattern quality of the $t$-th pattern candidate set $C$ is:
\begin{equation}
Pattern(C) = \frac{1}{|C|}\sum_{i=1}^{|C|}\text{min}\{\psi(C_i, S_{t_j})\}_{j=1}^{j=|S_t|}
\end{equation}
Where $\psi$ is the Cosine similarity function to evaluate the similarity between a seed and a generated address.

\begin{table}[tbp]
\caption{The Detail of Two Datasets Used in the Paper.}
\begin{center}
\begin{tabular}{llll}
\toprule
Dataset & Description & Period & \#Seeds\\
\midrule
IPv6 Hitlist & Active addresses & June 27, 2020 & 610.9k\\
 & Source addresses &&100.0k\\ 
 & Aliased prefixes &&516.1k\\  
\midrule
CERN IPv6 2018 & Active addresses & March - July 2018 & 90.1k\\
\bottomrule
\end{tabular}
\label{tab2}
\end{center}
\end{table}

\begin{table*}[htbp]
\caption{Results of 6GAN's Three Seed Classification Methods with 50k Budget on Each Pattern and 50k Total Seeds.}
\begin{center}
\begin{tabular}{llllllllllll}
\toprule
\multicolumn{3}{l}{RFC Based}&&\multicolumn{3}{l}{Entropy Clustering}&&\multicolumn{3}{l}{IPv62Vec}&\\
\cline{1-3}
\cline{5-7}
\cline{9-11}
Pattern & \#Seeds & \#Targets && Pattern & \#Seeds & \#Targets && Pattern & \#Seeds & \#Targets & Common Seed Example\\
\midrule
Embedded-IPv4&10.1k&13.0k&&$E_1$&9.1k&1.7k&&$I_1$&3.8k&3.7k&2001:db8:ff01:2::c8c3:8c07\\
Embedded-port&8.9k&3.7k&&$E_2$&9.5k&1.4k&&$I_2$&8.0k&6.9k&2001:db8::80\\
IEEE-derived&5.0k&4.0k&&$E_3$&5.2k&2.9k&&$I_3$&17.1k&5.8k&2001:db8:900::21e:67ff:fe31:4cdf\\
Low-byte&4.8k&1.5k&&$E_4$&16.0k&10.1k&&$I_4$&20.5k&13.5k&2001:db8:100:100::1\\
Pattern-bytes&15.0k&21.5k&&$E_5$&3.3k&2.4k&&$I_5$&0.3k&0.1k&2001:db8:ff01:4:face:b00c::a7\\
Randomized&6.2k&10.2k&&$E_6$&6.9k&25.5k&&$I_6$&0.3k&0.1k&2001:db8:8:68d3:b791:8741:c127:a75\\
\midrule
Total&50.0k&53.9k&&&50.0k&44.0k&&&50.0k&30.1k&\\
\bottomrule
\end{tabular}
\label{tab3}
\end{center}
\end{table*}

\item \textbf{Novelty quality.} The novelty quality metric can measure the difference between the candidate set $C$ and the seed set $S$ to indicate the algorithmic ability to generate new address sequences. Given the seed set $S$, the novelty quality of a candidate set $C$ is:
\begin{equation}
Novelty(C) = \frac{e}{|C|}\sum_{i=1}^{|C|}(1 - \text{max}\{\varphi(C_i, S_j)\}_{j=1}^{j=|S|})
\end{equation}
Where $\varphi$ is the Jaccard similarity function to measure whether the sampled nybble set is the same as the seed. $e = 100$ is a constant to help enlarge the metric value.
\item \textbf{Diversity quality.} The diversity quality metric assessment whether a candidate set $C$ is a diverse address set, which contains a variety of address sequences. The diversity quality of a candidate set $C$ is:
\begin{equation}
Diversity(C) = \frac{e}{|C|}\sum_{i=1}^{|C|}(1 - \text{max}\{\varphi(C_i, C_j)\}_{j=1}^{j=|C|, j \neq i})
\end{equation}
\item \textbf{Hit rate.} The hit rate refers to the proportion of active addresses in the candidate set, which could verify the model's learning ability. Assume $T$ is the real active target set in the IPv6 space and $T_a$ is the real aliased addresses set, the hit rate of a candidate set $C$ is: 
\begin{equation}
Hit(C) = \frac{|C\cap T - C\cap T_a|}{|C|}
\end{equation}
Where $C\cap T$ is the scanning result using Zmapv6 \cite{durumeric2013zmap}. We use known aliased prefix set to approximate $T_a$.
\item \textbf{Generation rate.} The generation rate is the proportion of the active addresses in the candidate set that are not in the seed set. The metric highlights the model's generation ability in the IPv6 space. Given the seed set $S$, assume the real active target set $T$ and real aliased addresses set $T_a$, the generation rate of a candidate set $C$ is:
\begin{equation}
Generation(C) = \frac{|C\cap T - C\cap T_a - C\cap S|}{|C|}
\end{equation}
\end{itemize}

\begin{figure}[htbp]
\centerline{\includegraphics[width=0.55\textwidth]{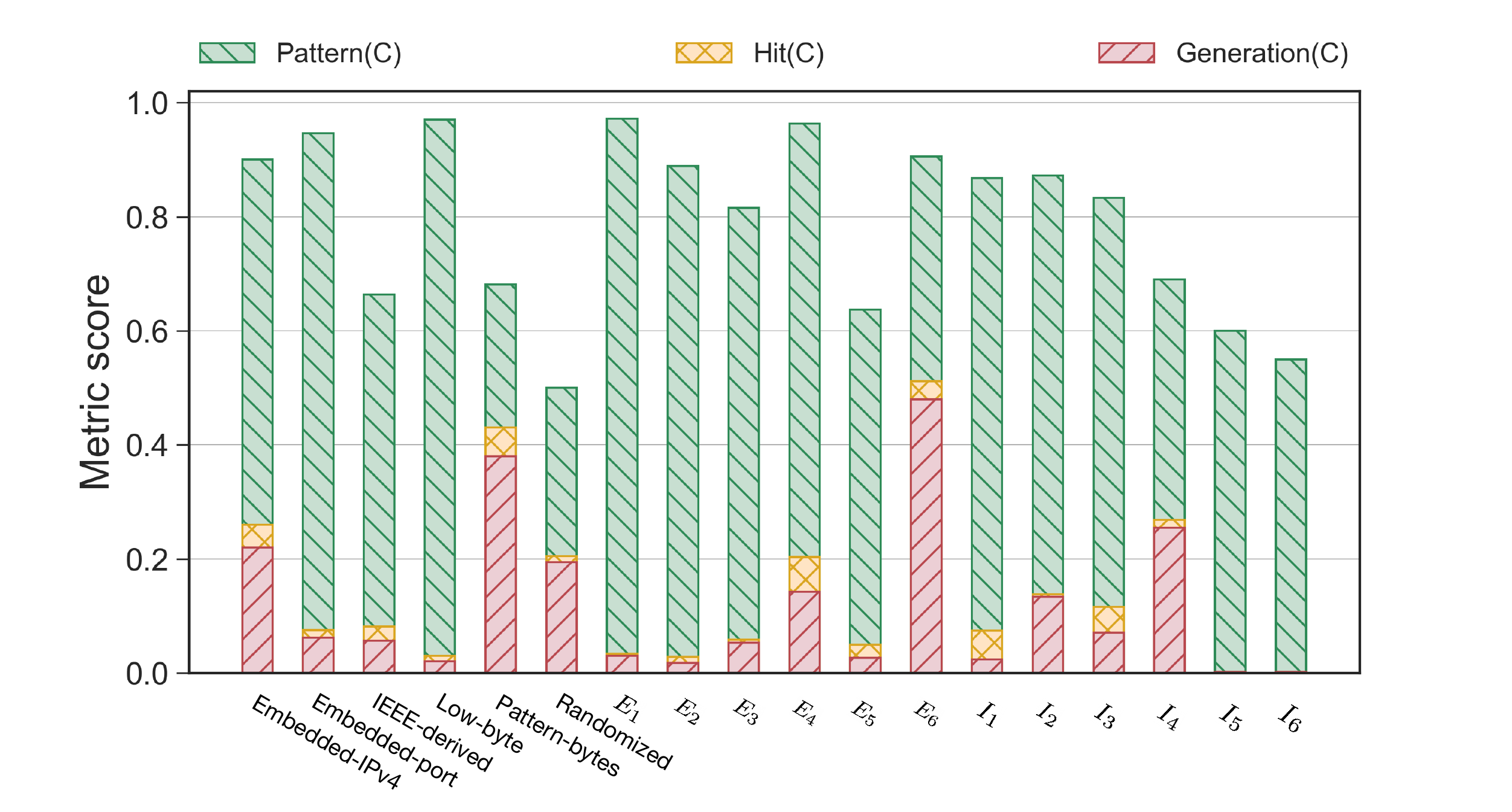}}
\caption{Three metric scores of all pattern candidates generated by generators with 50k budget on each pattern and 50k total seeds of IPv6 Hitlist dataset.}
\label{fig2}
\end{figure}


\textbf{4) Model Setting.} 6GAN is required to first transform nybble word in an address to a word embedding with the dimension size of 200. Word embeddings are randomly initialized while training. The generators are set as single-layer LSTM with the hidden dimension size of 200 and the max sample length of 32 nybbles. The discriminator is set as CNN with multiple kernel sizes from 1 to 16 and is added a highway architecture \cite{srivastava2015training} based on the pooled feature maps.


In adversarial training, the $\alpha$ is 0.9 to control the ratio of the discriminator reward and the alias detector reward. The reward strength of the alias detector $\lambda$ is 10. The sampling times $N$ in Monte Carlo search is set as 15. In the pre-training step, we pre-train generators for 60 steps and pre-train the discriminator for 20 steps. When adversarial training, the number of g-steps is 5 and the number of d-steps is 1. The optimization algorithm is RMSProp.

\subsection{Pattern Target Generation}
To explore the influence of addressing pattern on target generation tasks and 6GAN's performance on each addressing pattern, we separately use three seed classification methods to classify the seed set to determine the $k$ possible patterns in the IPv6 space. RFC Based could divide the seeds into 6 addressing patterns described in RFC 7707 \cite{chown2016network}. To keep the number of generation patterns consistent, we set the parameter $k$ to 6 in entropy clustering and adjust the parameters of DBSCAN \cite{ester1996density} to maintain 6-categories result in IPv62Vec. Table~\ref{tab3} shows the classification results and target generation results under the three classification methods of 6GAN on active addresses in the IPv6 Hitlist dataset.

\begin{table}[tbp]
\caption{Results of Classification Modes with 50k Budget Allocation and 50k Total Seeds.}
\begin{center}
\begin{tabular}{llll}
\toprule
Seed Classification & Budget Allocation & \#Targets & Generation(C)\\
\midrule
None&1&0.5k&1.06\%\\
RFC Based&11:3:3:1:19:10&12.7k&25.43\%\\
Entropy Clustering&2:1:3:8:1:26&16.9k&33.82\%\\
IPv62Vec&13:70:40:141:1:1&9.1k&18.19\%\\
\bottomrule
\end{tabular}
\label{tab4}
\end{center}
\end{table}

There are common addresses between the classified patterns of the different seed classification methods, which indicates that these methods may find similar addressing patterns. However, the classification boundaries might be slightly different. Pattern-bytes pattern includes the most address in RFC Based method, while more addresses are classified into $E_4$ and $I_4$ pattern in Entropy Clustering and IPv62Vec. To exploit the pattern generation performance of 6GAN, we set a 50k budget for each pattern to uncover the active targets hidden in the pattern. Pattern-bytes, $E_6$, and $I_4$  enable 6GAN to reach the most hit among the three seed classification methods.

6GAN forces generators to simulate the addressing patterns through adversarial training to achieve the purpose of deceiving the discriminator. In Figure \ref{fig2}, we use the pattern quality and hit rate of the candidate sets to illustrate the learning ability of 6GAN for each addressing pattern. The experiment shows that 6GAN has a strong ability to imitate most patterns. The low score of the patterns like $I_5$ and $I_6$ may be caused by the small size of the classified seed set. In addition, we calculate the generation rate of each pattern candidate set to reveal the active user distribution in the addressing patterns and help guide 6GAN's budget allocation. Given the generation rates of $k$ patterns $(r_1, ..., r_i, ..., r_k)$ and the total budget $|C|$, the allocated budget of $i$-th pattern is $|C_i|$ in optimized 6GAN: 
\begin{equation}
|C_i| = \frac{r_i}{\sum_{j=1}^kr_j}\times |C|
\end{equation}
Therefore, the budget allocation of 6GAN could be represented as $(|C_1|: ...:|C_k|)$.

Table \ref{tab4} shows the performance of 6GAN with budget allocation under the three seed classification methods. We compared the none classification method case which uses one single generator to learn the seed set. The 18.2-33.8 times discovered targets compared to none seed classification highlight the significance of pattern discovery. Under the three classification modes with budget allocation, Entropy Clustering achieves the best performance on the IPv6 Hitlist dataset contributed by the high generation rate of $E_6$. We strongly recommend readers to perform the budget allocation work to help 6GAN reach the best results on target generation.

\subsection{Pattern Discrimination}
Through adversarial training with multiple generators, 6GAN's discriminator can be optimized to achieve pattern discrimination. To explore the utility of the discriminator, for a case study, we use 50k seeds in the IPv6 Hitlist dataset to train 6GAN with RFC Based classification mode and label other 50k addresses in the dataset with RFC Based classification method to test the trained discriminator. Table \ref{tab5} indicates the pattern discrimination ability of 6GAN's discriminator. In most of the addressing patterns, the discriminator keeps an outstanding identification performance. Pattern-bytes obtains a lower accuracy than other pattern types because of mixing multiple unobvious patterns. The overall accuracy of the discriminator reaches 0.966 scores in the discrimination evaluation of the 6 pattern types. We believe that the trained 6GAN discriminator possess sufficient capacity to recognize addressing patterns in the network space, which allows researchers to fast infer host attributes to help wide-range network assessments. 





\subsection{Performance of Alias Detection}
In the architecture of 6GAN, the de-aliasing work is accomplished through the supervision of an alias detector when model training. To indicate the alias detection performance, we operate the alias detection condition and separately train 6GAN with 50k active addresses and 50k source addresses in the IPv6 Hitlist dataset, where the active addresses are all non-aliased addresses while the source addresses contain 7.9k aliased addresses and 42.1k non-aliased addresses. Table \ref{tab6} shows the experimental results of alias detection. 

Even if 6GAN is fully trained with the active address set, the sampling process of the generative model still keeps the opportunity to recombine the aliased prefix to generate the aliased address. By increasing the reward guidance for alias detection, 6GAN's generator could intelligently avoid exploiting alias regions to achieve effective alias detecting performance. When forced to learn in the dataset containing considerable aliased addresses, 6GAN's alias detector can still greatly reduce the generation of aliased addresses. This design effectively prevents wasting the budget of the candidate set to generate aliased addresses, thus enabling 6GAN to generate high-quality candidate sets.

\begin{table}[tbp]
\caption{Case Study of Pattern Discrimination Ability of 6GAN's Discriminator on RFC Based Addressing Patterns.}
\begin{center}
\begin{tabular}{llllll}
\toprule
Category & \#Labels &\# Preds& \#Hits &\#Errors& Accuracy\\
\midrule
Embedded-IPv4&4.38k&4.54k&4.17k&0.37k&0.954\\
Embedded-port&0.57k&0.52k&0.50k&0.02k&0.898\\
IEEE-derived&3.19k&3.37k&3.18k&0.19k&0.998\\
Low-byte&12.82k&12.04k&11.93k&0.11k&0.931\\
Pattern-bytes&0.73k&1.49k&0.51k&0.98k&0.701\\
Randomized&28.31k&28.04k&28.02k&0.02k&0.990\\
\midrule
Total&50.00k&50.00k&48.31k&1.69k&0.966\\
\bottomrule
\end{tabular}
\label{tab5}
\end{center}
\end{table}

\begin{table}[tbp]
\caption{Ablation Study on the Alias Detection with 50k Budget and Active / Source Seeds in the IPv6 Hitlist Dataset.}
\begin{center}
\begin{tabular}{llll}
\toprule
Seed set & Alias Detection & \#Aliased Targets & Percentage\\
\midrule
Active addresses&W/o&0.01k&0.02\%\\
Active addresses&W/&0.00k&0.00\%\\
Source addresses&W/o&6.91k&13.82\%\\
Source addresses&W/&0.01k&0.02\%\\
\bottomrule
\end{tabular}
\label{tab6}
\end{center}
\end{table}

\subsection{Quality of Generated Addresses}
While in this paper we have described the effectiveness of each component in the 6GAN architecture in detail, as a target generation approach, we still want to know the superiority of 6GAN compared to other work. Since a variety of concrete instantiation has been successful on IPv6 target generation, previously measuring only on the target quantity under a limited candidate set could not provide a comprehensive evaluation of these works. We incorporate the quality of the candidate set into the perspective of judgment and present multiple metrics to evaluate the quality of the generated addresses, which could indicate the excellent performance of 6GAN compared to prior studies.

\begin{figure*}[htbp]
\centering
\subfigure[IPv6 Hitlist\quad\quad\quad]{       
\label{1}
\begin{minipage}[t]{0.48\linewidth}
\centering
\includegraphics[width=9.6cm]{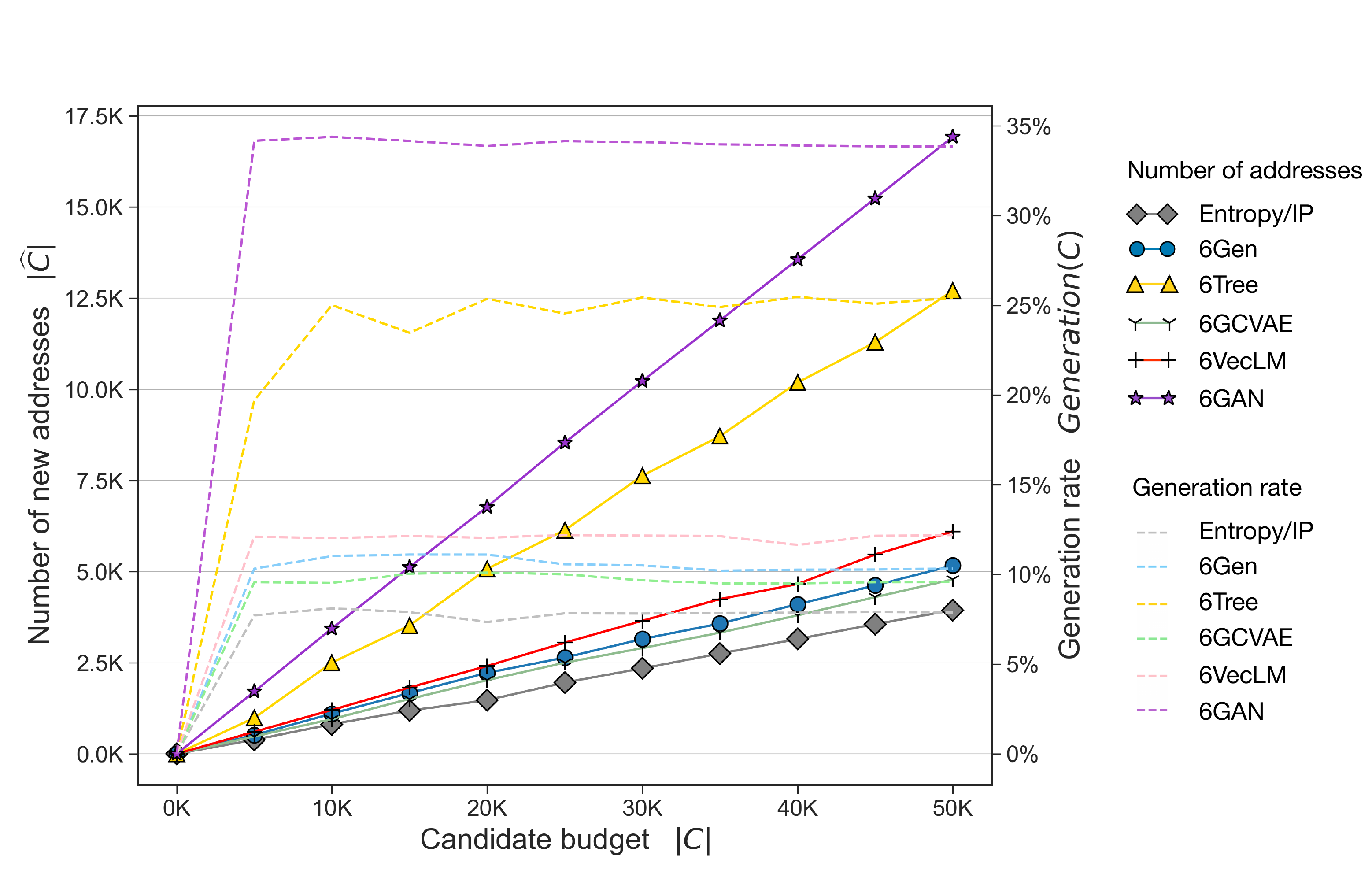}
\end{minipage}
}
\subfigure[CERN IPv6 2018]{ 
\label{2}
\begin{minipage}[t]{0.48\linewidth}
\centering
\includegraphics[width=7.7cm]{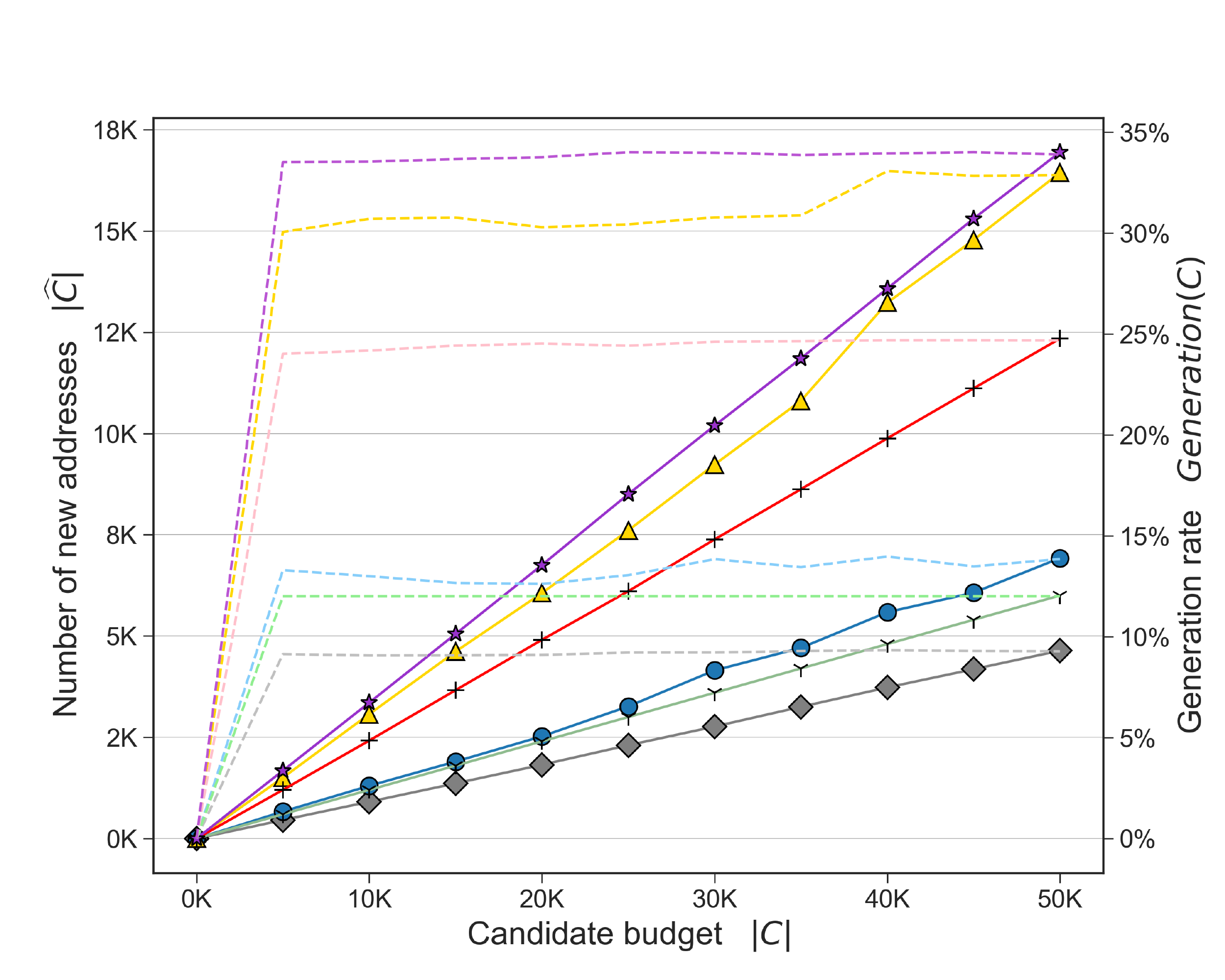}
\end{minipage}
}
\centering
\caption{Comparison of discovered addresses quantity $|\widehat{C}|$ and generation rate Generation(C) under different sizes of the candidate budget $|C|$ on the two datasets. 6GAN reaches the best performance through training with Entropy Clustering mode and setting 6 generators.}
\label{fig3}
\end{figure*}

\textbf{1) Baselines.} The baselines in our experiments for comparison mainly contain traditional design algorithms and deep learning approaches.
\begin{itemize}
\item \textbf{Traditional Design Algorithms.} Entropy/IP \cite{foremski2016entropy}, 6Gen \cite{murdock2017target}, and 6Tree \cite{liu20196tree} are the prior paradigms of the traditional design algorithm that have shown significant gains in target generation. Entropy/IP and 6Tree employ entropy information to construct address segments or space trees used to explore active targets, while 6Gen uses algorithmic analysis to uncover active clusters. In this paper, we use Entropy/IP and 6Tree's open-source code and reconstruct 6Gen according to the author's algorithm description to achieve the baselines of these works.
\item \textbf{Deep Learning Approaches.} 6GCVAE \cite{cui20206gcvae} and 6VecLM \cite{cui20206veclm} are the recent explorations of the target generation through deep learning. 6GCVAE is the generative model using gated convolutional Variational Autoencoder (VAE) \cite{kingma2013auto} architecture, while 6VecLM leverages Transformer \cite{vaswani2017attention} and softmax temperature \cite{muller2019does} to build the IPv6 language model for predicting address sequences. Since 6GAN is the same deep learning architecture, we use 6GCVAE and 6VecLM implemented through open-source code as the baselines for comparison to highlight the superiority of 6GAN's neural network architecture.
\end{itemize}

\begin{table}[tbp]
\caption{Candidates Quality of Target Generation Algorithms with 50k Budget and 50k Seeds.}
\begin{center}
\begin{tabular}{lllll}
\toprule
Approach & Novelty(C) & Diversity(C) & Hit(C) & Generation(C)\\
\midrule
Entropy/IP&12.37&6.80&12.03\%&7.88\%\\
6Gen&11.09&2.05&14.81\%&10.33\%\\
6Tree&11.16&2.06&24.40\%&24.39\%\\
6GCVAE&12.00&\textbf{7.66}&13.61\%&9.50\%\\
6VecLM&12.35&6.03&33.16\%&12.20\%\\
\textbf{6GAN}&\textbf{12.75}&4.73&\textbf{36.05\%}&\textbf{33.21\%}\\
\bottomrule
\end{tabular}
\label{tab7}
\end{center}
\end{table}

\textbf{2) Candidates Quality.} To comprehensively evaluate the performance of target generation algorithms, we adopt four metrics to measure the quality of the candidate sets generated by all approaches. Table \ref{tab7} shows the evaluation results of all target generation algorithms on the IPv6 Hitlist dataset. As traditional design algorithms, 6Gen and 6Tree achieve a lower novelty and diversity quality because they enumerate the addresses in the active regions, thus less changing the nybble set. Entropy/IP and 6GCVAE keep well exploiting the new addresses while they are rarely active targets. As deep learning architecture with generative adversarial nets, 6GAN could generate creative addresses with high novelty quality. However, 6GAN forces generators to learn the addressing patterns, the generated addresses keep similar sequences in each pattern type, thus obtaining a not high diversity quality score. We think the problem could be improved by setting more pattern types $k$ in seed classification for generator learning. Through a deep analysis in each addressing pattern, 6GAN outperforms all the baseline on the hit rate and generation rate in our experiments.

To explore the robustness of 6GAN to perform target generation under different datasets, we measure target quantity and generation rate under the different sizes of the budget with 50k total seeds in the IPv6 Hitlist and CERN IPv6 2018 datasets in Figure \ref{fig3}. Experimental results show that 6GAN could achieve a stable performance on both the two datasets. While comparing with the state-of-the-art target generation algorithm 6Tree, 6GAN could discover 1.03-1.33 times more active addresses than 6Tree in the limited budget. In addition, 6GAN's candidates keep different types of addressing patterns. The generators are controllable for researchers to generate specific pattern type addresses as long as training with the corresponding pattern seed data. We believe that this architecture implements a meaningful target generation work and could help more researchers chasing IPv6 addressing patterns.

\section{Conclusion}\label{sec6}
In this work, we explore the implementation of target generation algorithms to help address discovery in the global IPv6 address space. We present 6GAN, a generative adversarial nets architecture with reinforcement learning to achieve multi-pattern target generation.

6GAN uses multiple generators to learn the addressing patterns existing in IPv6 networks and optimizes them through the rewards of the discriminator and alias detector. 6GAN's generator can generate diversified addresses of different addressing pattern types. The discriminator employs a multi-class objective to recognize multiple pattern categories. 6GAN's alias detector provides rewards to avoid exploiting aliased prefixes and help 6GAN generate non-aliased candidate sets.  Experiments indicate that 6GAN outperformed the state-of-the-art target generation algorithms in multiple metrics. 

\section*{Acknowledgment}
This work is supported by The National Key Research and Development Program of China (No. 2020YFE0200500, No. 2018YFB1800200 and No.2020YFB1006100) and Key research and Development Program for Guangdong Province under grant No. 2019B010137003.

\balance
\bibliographystyle{IEEEtran}
\bibliography{IEEEabrv,mybibfile}

\end{document}